\documentstyle[11pt,epsf]{article}
\def\plotone#1{\centering \leavevmode
\epsfxsize=\textwidth \epsfbox{#1}}

\newcommand {\src} {V1432~Aql}

\newcommand {\etal} {{\sl et al.\/}\ }

\newcommand {\sqig} {$\sim$}

\newcommand {\rosat} {{\it ROSAT}}
\newcommand {\exosat} {{\it EXOSAT}}

\begin{document}

\
\vspace*{0.5 cm}

\thispagestyle{empty}

\begin{centering}
{\Large\bf An alternative model of \\
of the magnetic cataclysmic variable V1432 Aquilae \\
(=RX~J1940.1$-$1025)}

\vspace*{1.0 cm}

{\large Koji Mukai\footnote{Also Universities Space Research Association;
		e-mail address: mukai@lheavx.gsfc.nasa.gov} }

\vspace*{1.0 cm}

\end{centering}

Laboratory for High Energy Astrophysics,
NASA Goddard Space Flight Center, Greenbelt, Maryland 20771, USA.

\vspace*{1.5 cm}

\begin{quote}

{\bf Abstract:}

\src\ is currently considered to be an asynchronous AM Her type system,
with an orbital period of 12116.3 s and a spin period of 12150 s.  I
present an alternative model in which \src\ is an intermediate polar
with disk overflow or diskless accretion geometry, with a spin period
near 4040 s.  I argue that published data are insufficient to distinguish
between the two models; instead, I provide a series of predictions of the
two models that can be tested against future observations.

\end{quote}

\vspace*{1.5 cm}

\vspace*{1.5 cm}

\begin{centering}

Accepted for publication in the {\sl Astrophysical Journal}

\end{centering}

\newpage
\setcounter{page}{1}

\section{Introduction: the current model of \src}

NGC~6814 became one of the most celebrated Seyfert galaxies,
with the discovery of the $\sim$12000s periodicity by Mittaz \&
Branduardi-Raymont (1989) in the \exosat\ ME data.
However, a \rosat\ observation revealed another bright X-ray source,
RX~J1940.1$-$1025, which turned out to be the origin of the \sqig 12000 s
periodicity: it now appears likely that past non-imaging X-ray observations
of `NGC~6814' are in fact largely that of this newly discovered source.
Subsequently, RX~J1940.1$-$1025 was identified with a cataclysmic variable
(Watson et al 1995; see also Staubert et al
1994), and has recently been given the variable star designation \src.

A cataclysmic variable is an interacting binary in which the secondary
(mass donor) is a late type star on or near the main sequence, and the
primary (mass accretor) is a white dwarf.  Watson \etal
(1995) have adopted the AM Her sub-class for \src: in an AM Her system,
the magnetic field of its primary is strong enough to (i) prevent the
formation of an accretion disk; (ii) emit cyclotron radiation in the
optical thus making it polarized; and (iii) synchronize the spin to
the orbital periods (see Cropper 1990 for a review).

\src\ appeared to distinguish itself among AM Her type systems with
its strong hard X-ray component and the extreme complexity of its
X-ray light curves.  In most AM Her type systems, the blackbody-like soft
X-ray component from the heated surface of the primary is substantially
stronger than the bremsstrahlung-like hard X-ray component from the
post-shock region at the base of the accretion column (see, e.g.,
Ramsay et al 1995).  In contrast, in intermediate polars
(IPs, also known as DQ Her stars; see Patterson 1994
for a review), which are defined as magnetic CVs in which the white dwarf
spin period is much shorter than the orbital period, the hard X-ray
component usually dominates over the soft: in this respect, \src\ is
more similar to the IPs, in particular the recently recognized subclass of
soft IPs (Haberl \& Motch 1995). 

In fact, extensive observations of \src\ by Patterson \etal
(1995) and also by Friedrich \etal (1996) have revealed
evidence for multiple periodicities, one at 12116.3 s and another at
around 12150 s.  They both interpret the former, defined by deep dips
seen both in the X-ray and in the optical, as the orbital period of the binary.
However, their interpretations differ on the nature of the dips: Patterson
\etal (1995) favor an eclipse by the secondary, while Friedrich
\etal favor absorption dip due to the accretion stream, as originally proposed
by Watson \etal (1995).  They do agree in interpreting the
longer period as the spin period of the primary.  They therefore grouped
\src\ with V1500~Cyg and BY~Cam as an asynchronous AM Her type
system\footnote{In these systems, the two periods are within a few percent
of each other, compared with IPs in which the orbital period is typically
\sqig 10 times the spin period.}, although in both V1500~Cyg and BY~Cam the
spin period is shorter, whereas in \src\ the reverse is claimed.

If we adopt the asynchronous AM Her model for \src, we are faced with
the following puzzles.

One is the extremely complex shapes of the folded X-ray light curves
in \src\ (e.t., Done \etal 1992; Madejski \etal
1993; Staubert \etal 1994; Watson \etal
1995).  A part of this complexity can be explained as
due to the asynchronous nature of \src, if the X-ray observations spanned
many days.  However, in relatively short observations, which should not
suffer from this problem, the light curves are still complex and unique
among AM Her type systems.  In particular, the appearance of two distinct
bright phases separated by faint phases of unequal durations in \rosat\
data is unprecedented among AM Hers.  Additionally, an asynchronous AM Her
is expected to change its light curves drastically as a function of the
supercycle (in this case $(P_{orb}^{-1} -P_{spin}^{-1})^{-1}$, where
$P_{orb}$ is the orbital period and $P_{spin}$ is the spin period),
as accretion switches from one pole to the other.  The flaring vs pulsing
behaviors seen in BY~Cam in the Ginga observation (Ishida \etal
1991) may be a manifestation of pole-switching; no such change of states
has been detected in \src\ to date.

The other is the question of how the spin period of \src\ has become longer
than its orbital period.  The usual scenario of spin evolution of magnetic
CVs is that they evolve from IPs ($P_{spin} \ll P_{orb}$) to AM Hers
($P_{spin} = P_{orb}$); the timescale of the spin evolution is shorter
than that of the orbital evolution, and so the models have not included
a system with $P_{spin} > P_{orb}$.  A classical nova outburst can
temporarily force AM Her systems off synchronism; however, the case
of V1500~Cyg (=Nova Cygni 1975) demonstrates that the spin period should
become shorter after a nova outburst.

In an effort to resolve these difficulties, I have re-investigated the X-ray
periodicities of \src\ using archival \rosat\ data (\S 2), and propose
an alternative model of \src\ (\S 3).

\section{Data analysis and results}

I have started with a list of all \rosat\ observations of the
NGC 6814/\src\ region of the sky in the public archive (see Table 1); all these
observations have been reported in the various references cited above.

\begin{table}
\begin{centering}
\caption{ROSAT Observations of the NGC 6814/\src\ Region}

\vspace{0.25 cm}

\begin{tabular}{ccccrl}
\hline\hline
Label & ROR & From & To & Exposure & Note \\
\hline
 & 700782 & 1992 Apr 29 & 1992 Apr 29 & 8820s & Severe strut obscuration; not used \\
(a) & 701090 & 1992 Oct 08 & 1992 Oct 30 & 29684s & Oct 27 -- Oct 30 used \\
(b) & 700923 & 1993 Mar 31 & 1993 Apr 02 & 38244s & \\
(c) & & 1993 Oct 13 & 1993 Oct 18 & 31428s & 4 RORs combined \\
\hline
\end{tabular}

\end{centering}

\end{table}

The first observation was short and affected strongly by the unfortunate
positioning of \src, so it will not be discussed further.  In the
1992 Oct and 1993 Mar/Apr, a relatively large extraction region was
necessary due to the off-axis location of \src; for these observations,
I have subtracted background light curves taken from nearby regions.
For the 1993 Oct observation, a small extraction region could be used
and the background was judged to be negligible.  Although the 1992 Oct
observation was obtained over a $\sim$22 day period, most of the data
were in fact obtained during the last 4 days; here I will present only
the light curves from the period 1992 Oct 27--30.

I have used PSPC channels 10--40 (\sqig 0.1--0.4 keV) as the soft band
and 50--200 (\sqig 0.5--2.0 keV) as the hard band.  These bands are
dominated by the soft and the hard spectral component, respectively.
I have analyzed them separately, since the soft and the hard components
are modulated differently in some magnetic CVs (see, e.g., Allan \etal
1996).

In Figure 1, I present the periodogram of \src\ using the definition of
Scargle (1982).  Note that the the complicated alias
patterns caused by the orbital period of the \rosat\ satellite
(\sqig 5760 s).  Dashed lines indicates period of 12120s, 6060s, and
4040s (at the resolution of these plots, 12116.3s and 12150s
would be indistinguishable).  It is interesting to note that the 12120 s
period is relatively weak in many of the datasets; two of the hard band
periodograms show a strong 6060 s period, while the 4040 s period is
almost always important.

\begin{figure}
\plotone{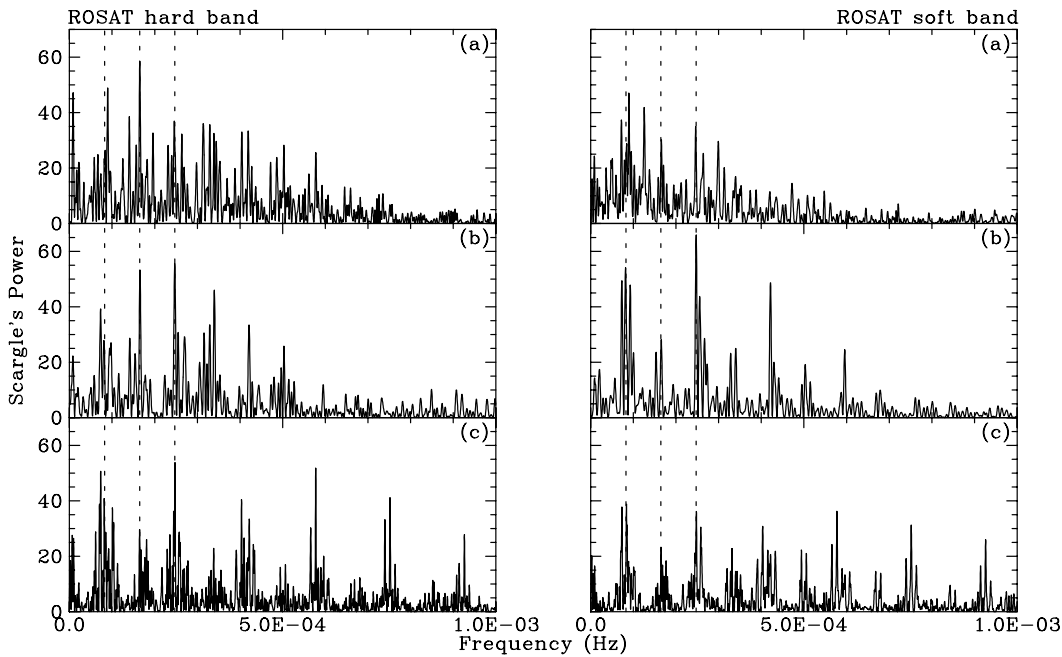}
\caption{The periodograms of 3 \rosat\ observations of \src\ 
(labelled according to the symbols used in Table 1), calculated using
the definition of Scargle (1982), in the soft and hard energy bands.
Three dashed lines indicate (from left to right) 12120 s, 6060 s, and
4040 s respectively; the frequency resolution of the \rosat\ data is
insufficient to distinguish 12120 and 12150 s periods. }
\end{figure}

To investigate this, further, I have folded these light curves on
a 4040 s trial periods; the results are displayed in Figure 2.
The error bars are the standard deviations, which may be more
appropriate than those derived by propagating errors, since \src\ is
intrinsically variable on various time scales.  It appears that folded
light curves of \src\ are far more readily interpretable if this is
the spin period; in particular, in the soft band, a simple bright phase/
faint phase structure common among many AM Hers can be used as basis for
interpretation.

\begin{figure}
\plotone{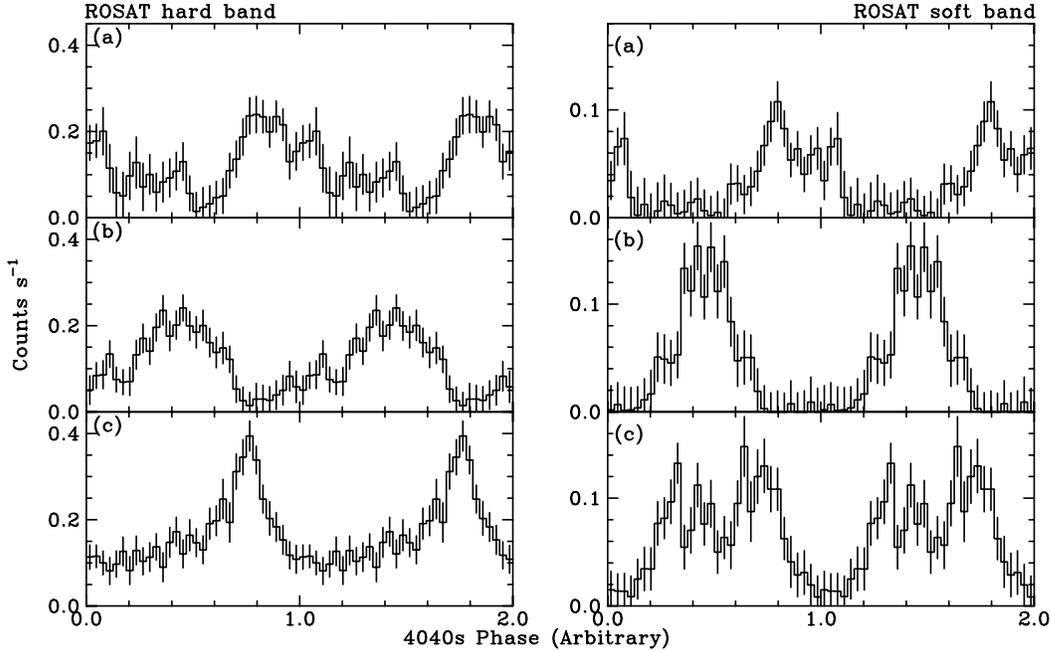}
\caption{The two-band \rosat\ light curves of \src,
folded on a 4040 s period with arbitrary origins.  Two cycles are
plotted in each panel for clarity. }
\end{figure}
 
\section{A alternative model: P$_{spin}$ \sqig 4040s}

In this model, the orbital period of \src\ remains 12116.3 s:
this value comes from the recurrence period of the eclipses,
also seen in the radial velocity curves of the narrow component
of optical emission lines.  In addition to the eclipse, both the
optical and X-ray light curves show a much broader orbital modulation
(an X-ray orbital minimum around phase 0.8 is commonly seen in many
high inclination IPs; Hellier \etal 1993).

No data on \src\ available to date appear to {\sl requires\/} a true X-ray
periodicity of $\sim$12150 s.  Many datasets do not have the frequency
resolution to distinguish between 12116.3 s and 12150 s period, and
$\sim$12000 s period from such data can be interpreted as an orbital
phenomenon.  When the detection of a 12150 s period is claimed, it is
often based on folding analysis, which cannot distinguish between a
12150 s period and a 6075=12150/2 or 4050=12150/3 s periods.  The apparent
detection of a second period at $\sim$12150 s in the X-rays and in the optical
can be explained in one of three ways:

\begin{enumerate}
\item $P_{spin} \sim$ 4050.0 s; in this case 12150 s is simply 3$P_{spin}$.
\item $P_{spin} \sim$ 4046.25 s; in this case, 12150 s is twice the sideband
	period $(P_{spin}^{-1} - P_{orb}^{-1})^{-1}$.
\item $P_{spin} \sim$ 4042.5 s; in this case, 12150 s is the other possible
	sideband period, $(P_{spin}^{-1} - 2P_{orb}^{-1})^{-1}$ (see, e.g.,
	Warner 1986).
\end{enumerate}

The latter two possibilities require that the accretion geometry be
diskless (see, e.g., Wynn \& King 1992), or of the
disk overflow type (Hellier 1993).  An X-ray periodicity at
the sideband period $(P_{spin}^{-1} - P_{orb}^{-1})^{-1}$ is a natural
consequence of these accretion geometries, whereas reprocessing can produce
both the first and the second sidebands in the optical (Warner
1986).  Given the strength of the 6075 s period in the extensive optical
photometry of Patterson \etal (1995), as well as in several
\rosat\ hard-band light curves, the $P_{spin} \sim$ 4046.25 s case appears
most promising.  For all three cases, the dips recurring at the 12116.3 s
must be due to an eclipse by the secondary, since the accretion
stream will not stay in a fixed point in the binary frame.

Normally, if the Fourier periodograms reveal a peak at an integer multiple
(12150 s) of the most obvious period (4040 s), then this is a conclusive
evidence that the former is the fundamental period.  However, this reasoning
does not apply here for three reasons.  First, the existence of the orbital
period at 12116.3 s complicates any attempt at interpreting the periodograms.
Second, the 12150 s is near the expected position of optical sidebands.
Third, for the X-rays, the alias pattern caused by the spacecraft orbital
period of \rosat\ (\sqig 5760 s) further complicates situation.  A folding
analysis that shows a peak at 12150 s (Done \etal 1992;
Madejski \etal 1993) does not necessarily mean that the
underlying period is 12150 s; a true period of 6075 or 4050 s would also
result in such a peak.

Adopting $P_{spin} \sim$ 4046.25 s as the most likely candidate, we can
summarize the new model as follows.  \src\ is an IP with a 4046.25 s spin
period and a 12116.3 s orbital period.  In the X-rays, we observe the spin
modulation due to our viewing geometry, sideband modulation at 6075 s due
to the pole-switching of the disk overflow or diskless accretion, and an
orbital modulation at 12116.3 s, including a sharp eclipse by the secondary
and a much broader minimum caused by an azimuthal structure on the partial
disk.  Such an orbital minimum at orbital phase 0.8 is commonly seen in IPs
and thought to be caused by the impact of the stream from the secondary on
the accretion disk (Hellier \etal 1993), although in this case
the broad minimum appears to start late ($\sim$0.9) and continue beyond the
eclipse ($\sim$0.1; see, e.g., Watson et al 1995).  The
optically thick, soft X-rays from the irradiated surface shows a clearer
spin modulation than the optically thin, hard X-rays from above the surface,
because only the former is modulated by the projection effect, whereas both
can be modulated by the absorption in the accretion stream.  The X-ray
modulations at 4046.25 s, 6075 s, and 12116.3 s periods combine to create
the impression that there is a 12150 s period.  In the optical, the strongest
signals are at the 6075 s sideband period, and at the orbital period; the spin
period and the second sideband period at 12184 s may also be present.

Given these considerations, there is as yet no conclusive observational
test that enables us to choose between the $P_{spin} \sim$ 12150 s and
$P_{spin} \sim$ 4040 s models.  However, with further observations of
the right type, a decisive test can be performed.

\section{Predictions of the two models}

In this section, I will present predictions of the two models
for three different timescales: Long-term, supercycle, and spin.

\subsection{Long-term period evolution}

In an equilibrium situation, there must be no net accretion
of angular momentum by the primary.  In the model originally
developed for X-ray pulsars by Ghosh \& Lamb
(1979) and subsequently
refined by these and other authors, material accretion torque
is balanced by the magnetic torque (stress between the magnetic
field line, anchored to the compact object, and the more slowly
rotating part of the disk).  In an asynchronous magnetic CV accreting without
a disk, interaction of the field with the accretion stream can
play a similar role.  In addition, synchronization torque arises
when the magnetic field of the primary can interact with the secondary.

It is important to note that, if the spin period of \src\ is 12150 s,
all these torques operate in concert (everything that the magnetic field can
interact with is moving faster, thus magnetic torque acts to spin up
the white dwarf; so does the accretion torque), unlike in proven
asynchronous systems such as V1500~Cyg where they compete with each other.
Watson (1995) estimates a luminosity of \src\ to be
\sqig 3 $\times 10^{32}$ ergs\,s$^{-1}$.  The inferred material
accretion torque alone leads to an order of magnitude estimate
of spin-up of \sqig 2 $\times 10^{-9}$, two orders of magnitudes
greater than commonly found in IPs.

V1500~Cyg, which is asynchronous ($P_{spin}$ is 2\% shorter then
P$_{orb}$) with a similar orbital period, has a synchronization timescale
of 150 yr.  This indicates that the synchronization torque is an order
of magnitude stronger than the accretion torque (Schmidt \& Stockman
1991).  If the same is true for \src, in which it acts
with the accretion torque, it should be synchronized with a time scale
perhaps much shorter than 150 yr.

Geckeler \& Staubert (1997) have used the `troughs' in
the optical and X-ray light curves, detected an O$-$C variation
on the supercycle, and in addition claimed a marginally significant
detection of a spin-up in \src.  However, the latter is dependent on the
choice of the initial spin period: they have refined the value of 12150.7 s
from the earlier work of Friedrich \etal (1996).  Note that
the analysis of Friedrich \etal was performed before the periodic O$-$C
variation was discovered, and should therefore be re-examined.  In Figure 3,
we plot a periodogram based on the optical `troughs' as reported in Table 1 of
Geckeler \& Staubert (this is a purer method than mixing optical and X-ray
troughs, as the O$-$C variations around the supercycle are expected differ
in details between the two bands even when the underlying cause is the same).
According to this analysis, a period of 12145.2s is somewhat preferred
although a 12150.1 s period is also possible.  Long-term O$-$C diagrams for
these two trial periods show that a significant $\dot P$ is only required
for the latter period, not the former.  I therefore conclude that it would
require a much more extensive data set than Geckeler \& Staubert
(1997) used uniquely to determine the precise value of the
`12150 s' period and to search for period changes.

\begin{figure}
\plotone{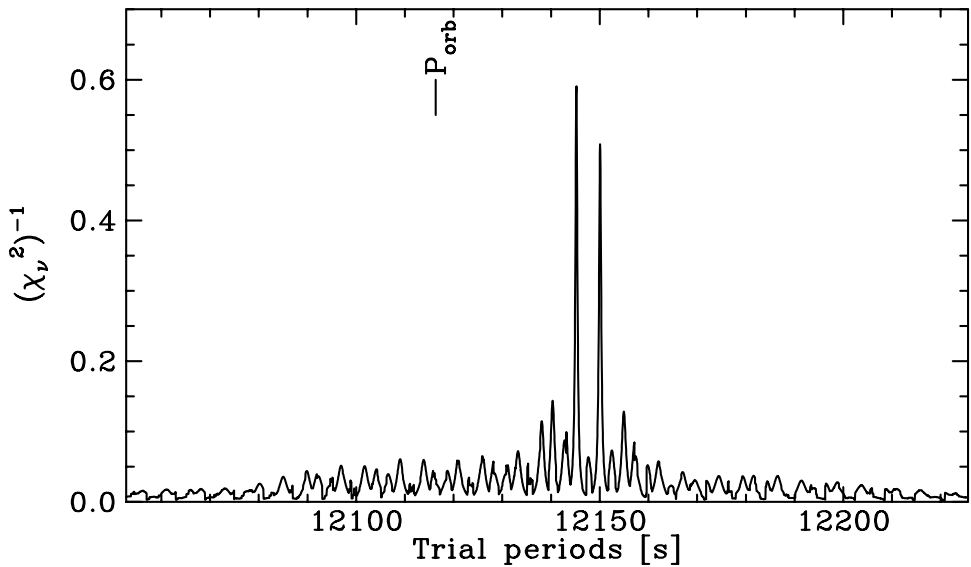}
\caption{The periodograms constructed from times of optical
`troughs', as reported by Geckeler \& Staubert (1997).  For each trial period,
the $\chi^2_\nu$ value was calculated for a linear ephemeris, and its
inverse plotted for clarity of presentation.  The highest is at 12145.2 s;
the second highest peak at 12150.1 s is the one adopted by Geckeler \& Staubert.
No significant peaks are seen around the 12116.3 s orbital period. }
\end{figure}

\subsection{Supercycle phenomena}

An orbital period of 12116.3 s and a spin period of 12150 s implies
a supercycle of \sqig 50 days (Patterson \etal 1995;
Friedrich \etal 1996).
The alignment of the spin and the orbital phenomena should change as
a function of the supercycle phase.  In particular, if the magnetic
field of the white dwarf is dominated by the dipole component, then
accretion should switch between the upper and lower poles twice per
supercycle (or at least switch from predominantly to the upper pole
to predominantly to the lower pole). 

There are several observable consequences of this pole-switching.
First and foremost, the spin light curves should show significant
changes, particularly in X-rays (optical light curves are `contaminated'
by the underlying stellar components and reprocessed light).  If
the orbital inclination is such that only the upper pole is visible
to us, then this would cause a pronounced low state lasting for $\sim$half
a supercycle.  On the other hand, if the orbital dips are caused by an eclipse
by the secondary, the orbital inclination is so high that both poles
are almost equally visible.  When pole-switching occurs, the peak due to
one pole should disappear to be replaced by the other pole, which will
lead to a 180$^\circ$ phase shift (although pole-switching may occur more
gradually, with periods when both poles are accreting, which would make the
phase shift less dramatic and harder to notice).  Neither has been reported
in \src\ so far.

If the orbital dip is caused by the accretion stream, it should disappear
while the lower pole accretes (accretion stream to the lower pole would
not cross our line of sight).  The dip timings should show a pronounced phase
jitter while the upper pole accretes, as the curvature of the accretion
stream changes.  However, the dip appears to be a persistent feature of
\src\ and tend to stay at the same orbital phase to within $\sim$100 s or
so (Watson \etal 1995).  If the dip is an eclipse by the
secondary, then a much subtler, but observable, phase jitter is expected,
as has been recently demonstrated for XY~Ari (Hellier 1997b).
Such considerations of the possible supercycle phenomena appear to argue
against the standard, asynchronous polar model of \src.

If \src\ is an IP, as I have proposed in this work, then its overall
appearance should remain essentially the same over the $\sim$50 day period.
However, there will be a subtle supercycle in the sense that the spin
phase at mid-dip will gradually change.  In general, if
$P_{orb} = (n+f)P_{spin}$, where $n$ is an integer and $f$ is a real number
in the range $-0.5\leq f\le0.5$, then the spin phase at a given orbital
phase changes by $f$ every spin cycle, creating a ``supercycle'' of
$P_{orb}/f$.  Since most IPs do not have a sharp orbital phase marker at which
to measure the spin phase, this supercycle usually is not noticeable.
However, \src\ has the dip, which we argue is an eclipse by the
secondary, and spin phase at eclipse can be measured (Hellier
1997b).  An extensive dip timing campaign over
a $>$50 day period may enable us to determine the supercycle of \src,
and thus provide a strong discriminator of the true spin period.
The length of the supercycle depends on the exact spin period:
Possible values are 50 d ($P_{spin}$ = 4042.5 s), 25 d ($P_{spin}$ =
4046.25), or 17 d ($P_{spin}$ = 4050 s).  If the long period is 12145 s,
rather than 12150 s as assumed above, then the true spin period would be
shorter and the supercycle longer.

\subsection{Spin phenomena}

So far, the analyses of time-resolved optical spectroscopy have
concentrated on the narrow component of the emission lines,
believed to originate on the heated face of the secondary (e.g.,
Friedrich \etal 1996).  It has been conclusively
established that many IPs show radial velocity variations at the
spin period, analyzed e.g., in the form of V/R ratio (see, for
example, Hellier (1997a) and references therein).
A similar study for \src\ would be most useful; a discovery of
a periodic V/R modulation at the 4040 s period would be a conclusive
evidence for the new model.

Optical polarimetry is another powerful method that is capable of
constraining the geometry.  If \src\ is an asynchronous polar, then
each night's data (particularly the position angles) can be analyzed
separately to derive binary inclination and the instantaneous co-latitude
of the cyclotron emitting region.  If the IP model is correct, on the
other hand, the polarization will be modulated on the spin period, the
sideband period, or both, rather than at the 12150 s period.

Time resolved X-ray spectroscopy may also provide a unique signature
of the spin period: the reflection of hard X-rays off the white dwarf
surface produces a characteristic spectral component, including the
fluorescent Fe line at 6.4 keV (Done \etal 1994).
This component should be modulated at the spin period of the white dwarf,
as the viewing geometry of the white dwarf surface changes.  Detection of
a \sqig 4040 s periodicity in the 6.4 keV line will also be a conclusive
evidence for the new model.

\section{Conclusions}

I have shown that a scenario exists in which the spin period of
\src\ is \sqig 4040 s, not 12150 s.  Since current data are
insufficient to discriminate between the two models, I have
provided a set of predictions that can be tested with future
observations.  It is important to do so: if the correct value
of $P_{spin}$ turns out to be 12150 s, this may indicate an
unexpected new twist in the physics of accretion torque.

On the other hand, if the new model is correct, it would be the second
case of near commensurate ratio of orbital and spin periods (the case
of EX~Hya is well documented; see, for example, Patterson
1994 and references therein).  The spin orbit resonance for the planet
Mercury (Klassen 1976) is well known; it may be that
IPs that are evolving towards synchronism may get trapped in a resonance.
However, it would be wise to await the confirmation or otherwise of
the new model for \src\ before making any further speculations.

\section*{Acknowledgments}

I thank Dr. Chris Done for stimulating conversations and
constant encouragement, and Dr. Greg Madejski for constructive
comments on an earlier draft.  I also thank an anonymous referee
for useful comments.

\newpage

\section*{References}

\begin{list}{}{\setlength{\leftmargin}{0.75 in}
	\setlength{\itemindent}{-0.75 in}
	\setlength{\parsep}{0.0 cm}
	\setlength{\itemsep}{0.0 cm}}

\item Allan, A., Horne, K., Hellier, C., Mukai, K.,
	Barwig, H., Bennie, P.~J. \& Hilditch, R.~W. 1996, MNRAS, 279, 1345.
\item Done, C., Madejski, G.M., Mushotzky, R.F., Turner, T.J.,
	Koyama, K. \& Kunieda, H. 1992, ApJ, 400, 138.
\item Done, C., Osborne, J.P. \& Beardmore, A.P. 1995,
	MNRAS, 276, 483.
\item Friedrich, S., Staubert, R., Lamer, G., K\"onig, M.,
	Geckeler, R., B\"assgen, M., Kollatschny, W., \"Ostreicher, R.,
	James, S.D., \& Sood, R.K. 1996, A\&A, 306, 860.
\item Geckeler, R.D. \& Staubert, R. 1997, A\&A, in press.
\item Ghosh, P. \& Lamb, F.K. 1979, ApJ, 234, 296.
\item Haberl, F. \& Motch, C. 1995, A\&A, 297, L37.
\item Hellier, C., 1993, MNRAS, 265, L35.
\item Hellier, C. 1997a, MNRAS, 288, 817.
\item Hellier, C. 1997b, MNRAS, in press.
\item Hellier, C., Garlick, M.A. \& Mason, K.O. 1993, MNRAS, 260, 299.
\item Ishida, M., Silber, A., Bradt, H.V., Remillard, R.A.,
	Makishima, K. \& Ohashi, T. 1991, ApJ, 367, 270.
\item Klassen, K.P. 1976, Icarus, 28, 469.
\item Madejski, G.M., Done, C., Turner, T.J., Mushotzky, R.F.,
	Serlemitsos, P., Fiore, F., Sikora, M., \& Begelman, M.C. 1993,
	Nature, 365, 626.
\item Mittaz, J.P.D. \& Branduardi-Raymont, G. 1989, MNRAS, 238, 1029.
\item Patterson, J. 1994, PASP, 106, 269.
\item Patterson, J., Skillman, D.R., Thorstensen, J., \&
	Hellier, C. 1995, PASP, 107, 307.
\item Ramsay, G., Mason, K.O., Cropper, M., Watson, M.G. \& Clayton,
	K.L. 1993, MNRAS, 270, 692.
\item Scargle, J.D. 1982, ApJ 263, 835.
\item Schmidt, G.D. \& Stockman, H.S. 1991, ApJ, 371, 749.
\item Staubert, R., K\"onig, M., Friedrich, S., Lamer, G.,
	Sood, R.K., James, S.D., \& Sharma, D.P. 1994, A\&A, 288, 513.
\item Warner, B. 1986, MNRAS, 219, 347.
\item Watson, M.G. 1995, in ``Cape Workshop on Magnetic
	Cataclysmic Variables'', eds. D.A.H. Buckley \& B. Warner, ASP
	Conference Series 85, 259.
\item Watson, M.G., Rosen, S.R., O'Donoghue, D., BUckley, D.,
	Warner, B., Hellier, C., Ramseyer, T., Done, C., \& Madejski, G.
	1995, MNRAS, 273, 681.
\item Wynn, G.A. \& King, A.R. 1992, MNRAS, 255, 83.

\end{list}

\end{document}